\documentclass[a4paper]{article}

\usepackage{ISCSLP_v2}

\usepackage{enumitem}
\graphicspath{{./pdf/}}
\DeclareGraphicsExtensions{.pdf}
\usepackage[caption=false,font=footnotesize]{subfig}

\title{Exploring a Unified Attention-Based Pooling Framework \\ for Speaker Verification}
\name{Yi Liu$^1$, Liang He$^1$, Weiwei Liu$^2$, Jia Liu$^1$}
\address{
  $^1$Tsinghua National Laboratory for Information Science and Technology, \\
  Department of Electronic Engineering, Tsinghua University, Beijing 100084, China \\
  $^2$62315 Unit, Chinese People's Liberation Army, Beijing 100842, China}
\email{liu-yi15@mails.tsinghua.edu.cn, \{heliang, liuj\}@tsinghua.edu.cn, liu-ww10@hotmail.com}

\begin{document}

\maketitle
\begin{abstract}
  The pooling layer is an essential component in the neural network based speaker verification.
  Most of the current networks in speaker verification use average pooling to derive the utterance-level speaker representations.
  Average pooling takes every frame as equally important, which is suboptimal since the speaker-discriminant power is different between speech segments.
  In this paper, we present a unified attention-based pooling framework and combine it with the multi-head attention.
  Experiments on the Fisher and NIST SRE 2010 dataset show that  involving outputs from lower layers to compute the attention weights can outperform average pooling and achieve better results than vanilla attention method. The multi-head attention further improves the performance.
\end{abstract}

\section{Introduction}

The key problem in speaker verification is how to deal with variable-length utterances. Traditional models, e.g. joint factor analysis (JFA) \cite{Kenny_JFA}, use Bayesian methods as a solution. Speakers are modeled by conditional probability distributions and the verification scores are log-likelihoods. Based on statistical modeling, the i-vector framework represents variable-length samples with low-dimensional fixed-length vectors. Various backend machine learning algorithms can be applied to improve the performance. However, the Bayesian methods imply some strong hypotheses that cannot be fulfilled in practical. Also, some latent variables cannot be well estimated during training.

Although neural networks have been introduced to speaker verification for many years, it is typically used to compute sufficient statistics for i-vector extraction \cite{Lei_dnn_ivec}.
It is not until recently that neural networks are used to extract speaker-discriminant vectors directly. Similar to i-vector, neural network based methods represent utterances with fixed-length vectors, which are also known as \emph{speaker embeddings}. Early attempts of speaker embedding include the d-vector, which was initially developed for text-dependent speaker verification \cite{Variani_dvector}. Li et al. revisited d-vector, and found the method performs well in text-independent tasks \cite{Li_dvector}. To obtain d-vector, features are extracted in a frame-wise style and all the features in one utterance are averaged. This is criticized that the network can only see local information while the speaker characteristics tend to be noisy at the frame time-scale.
Optimization on the whole utterance is a better choice. Long short term memory (LSTM) is able to handle sequential information and is introduced to speaker verification in \cite{Heigold_e2e}. The LSTM output at the last time stamp is connected to successive layers, since it is believed that LSTM encodes the entire sequence in the final output. However, some useful details are still missing in this case, especially when the speech duration is too long to remember.

To better capture speaker characteristics, temporal pooling is applied. The network is partitioned into frame- and utterance-level subnetworks by the pooling layer. Average pooling is the most popular pooling method to extract the utterance-level representations and is used in many publications \cite{Li_deep, Snyder_xvector}. Other methods, e.g. spatial pyramid pooling (SPP) \cite{Zhang_spp}, are also explored using different network architectures. Speaker embeddings have outperformed conventional i-vectors in many conditions and become a promising approach.

The implicit hypothesis behind the temporal average pooling is that every frame in the utterance is equally important. Although neural networks are powerful to extract useful information from the inputs, it is difficult to transform features from different phonetic spaces to the same speaker representation, not to mention the fact that a large portion of speech does not represent the distinctive characteristics of speakers \cite{phonetic_power}. As we know, different segments should make different contributions to the speaker embedding. Attention mechanism is proposed to cope with this problem. The attention mechanism is capable to model the sequence dependencies and is suitable for variable-length transduction modeling. It has been successfully applied in image caption \cite{attention_caption}, machine translation \cite{attention_translation, attention_Vaswani} and automatic speech recognition (ASR) \cite{attention_speech}, etc. In sequence-to-sequence speech recognition, the influence of the entire speech to a single recognized word is estimated by the attention weights. The attention mechanism has also been used in speaker recognition. In \cite{Zhang_attention_speaker}, the authors aggregated outputs from a convolutional neural network (CNN) with weights computed from both acoustic and phonetic information. Higher-order attentive pooling is presented in \cite{attentive_pooling}. The first and second-order moments are used to model the speaker characteristics. Several variants of attention layers are also reported in \cite{Chowdhury_attention_speaker}.

Most of the previous works only used the outputs of the last layer in the frame-level network to derive the attention weights. However, the representations extracted from other hidden layers provide different levels of feature abstractions, which may be a better source for the attention computation. In this paper, we first present a unified attention framework. We show that the variants in \cite{Chowdhury_attention_speaker} are special cases of the proposed framework. Multi-head attention is then introduced to further increase the modeling power. Experiments on Fisher and NIST SRE 2010 10s-10s evaluation show that the attention-based pooling framework achieves better results than the average pooling.

The organization of this paper is as follows. The network architecture and the baseline average pooling we use is briefly introduced in Section 2. Section 3 describes the proposed attention framework and multi-head attention. Our experimental setup and results are given in Section 4 and 5. The last section concludes the paper.

\section{Neural network speaker embedding}

\begin{figure}[t]
\centering
\includegraphics[width=0.85\linewidth]{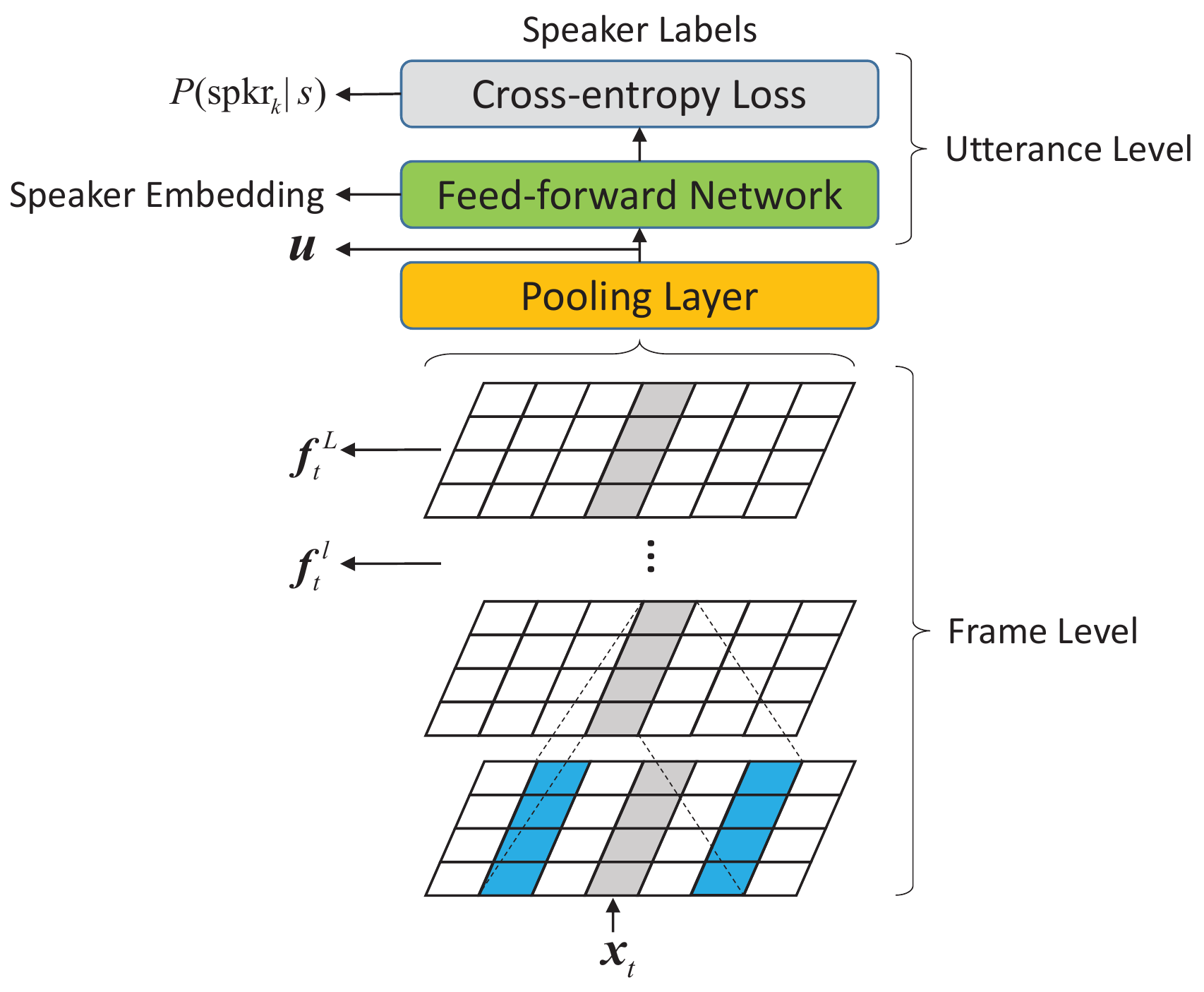}
\caption{The x-vector architecture for the speaker embedding extraction. The pooling layer splits the network into frame- and utterance-level networks. The superscript s is omitted for brevity.}
\label{fig:x_vector}
\end{figure}

In this paper, speaker embeddings are extracted by the x-vector architecture \cite{Snyder_xvector}. As shown in Fig. \ref{fig:x_vector}, the entire network is partitioned into frame and utterance levels. The input $\vec{x}_t^s$ is the feature of frame $t$ in utterance $s$. The frame-level network consists of $L$ hidden layers and the activation of each layer is denoted as $\vec{f}_t^{l,s}$.
A statistics average pooling component is applied to $\vec{f}_t^{L,s}$ and reduces them to an utterance representation $\vec{u}^s$. Fully-connected layers followed with a softmax layer are then used to predict the posterior of speaker $k$ with respect to utterance $s$. Given the parameters of the frame- and utterance-level networks, $\vec{\theta}_f$ and $\vec{\theta}_u$, the posterior $P(\text{spkr}_k|s)$ is expressed as:
\begin{equation}\label{xvector_frame}
  \vec{f}_t^{L,s} = \mathcal{F}(\vec{x}_t^s, \vec{\theta}_f)
\end{equation}
\begin{equation}\label{stat_pooling}
  \vec{u}^s = \mathcal{P}(\vec{f}_t^{L,s})
\end{equation}
\begin{equation}\label{xvector_utt}
  P(\text{spkr}_k|s) = \mathcal{F}(\vec{u}^s, \vec{\theta}_u)
\end{equation}
where $\mathcal{F}(\cdot)$, $\mathcal{P}(\cdot)$ represent the forward and pooling operations, respectively. We will omit the superscript $s$ for brevity hereinafter.

After training, the output of a hidden layer at the utterance-level network is extracted as the speaker embedding, termed the \emph{x-vector}. Linear discriminant analysis (LDA) and probabilistic linear discriminant analysis (PLDA) \cite{Garcia_plda} are further applied to compensate the session variabilities.

In this paper, we focus on the implementation of the pooling operation $\mathcal{P}(\cdot)$. Statistics average pooling treats all the inputs equally and calculates the mean and diagonal standard deviation. In the next section, we will show this can be replaced by the proposed multi-head attention-based pooling.

\section{Attention-based pooling}

\subsection{A unified framework}

Speech signal is complex and consists of many components. The speaker-discriminant power of speech segments can be affected by many factors, e.g., phonetic contents, acoustic environments and communication channels. 
Neural networks are capable to extract speaker information from raw features, but the process is hardly perfect. Some representations extracted from the frame-level network are more noisy and is less useful to model the speaker. In this paper, attention mechanism is used to automatically select the most speaker-discriminant segments. Motivated by \cite{attention_Vaswani}, we propose a unified attention framework for speaker verification.

Let us define a tuple $(\vec{v}_t, \vec{k}_t, \vec{q})$, where $\vec{v}_t$ is the \emph{value} sequence with $d_v$ dimensions and is the input of the attention layer, $\vec{k}_t$ is the \emph{key} sequence corresponding to $\vec{v}_t$ with $d_k$ dimensions, and $\vec{q}$ is a time-invariant \emph{query} with $d_q$ dimensions. The query maps the key sequence to the weights $\alpha_t$. The framework is expressed as
\begin{equation}\label{average_pooling_all}
  \text{Att}(\vec{v},\vec{k}, \vec{q}) = \left[ \hat{\vec{m}}', \hat{\vec{\sigma}}' \right]'
\end{equation}
\begin{equation}\label{attention_1}
  \hat{\vec{m}} = \sum_{t=1}^{T} \alpha_t \, \vec{v}_t
\end{equation}
\begin{equation}\label{attention_2}
  \hat{\vec{\sigma}} = \sqrt{\sum_{t=1}^{T} \alpha_t \, \text{diag}\left((\vec{v}_t-\hat{\vec{m}})(\vec{v}_t-\hat{\vec{m}})'\right)}
\end{equation}
\begin{equation}\label{attention_3}
  \alpha_t = \text{softmax}(\vec{q}' \cdot \mathcal{G}(\vec{k}_t, \vec{a}_t, \vec{\theta}_k))
\end{equation}
where $T$ is the length of the utterance, $\hat{\vec{m}}$ and $\hat{\vec{\sigma}}$ is the weighted mean and standard deviation and the softmax is performed among the time indices.

In speaker verification, the value is the output of the frame-level network, i.e. $\vec{v}_t = \vec{f}_t^{L}$. The query $\vec{q}$ is a parameter which can be learned by model optimization.
As in Eq. (\ref{attention_3}), the key is transformed by a compatibility function $\mathcal{G}(\cdot)$ and the transform parameter is $\vec{\theta}_k$.  An auxiliary feature $\vec{a}_t$ is also used in the transformation to provide extra information. If the auxiliary feature $\vec{a}_t$ is ignored and only consider $\vec{f}_t^{L}$ as the key, the model is called  \emph{self-attention}.

As the key is used to calculate the weights, it is required to indicate the potential speaker-discriminant power of the corresponding $
\vec{v}_t$. Empirically, we find that it is beneficial to replace the key with the activation of an intermediate layer. 
Previous study claimed that the phoneme information can guide the computation of the weights since different phonetic units have different speaker-discriminant power. During the network training, the activations $\vec{f}_t^{L}$ from the same speaker become less related to the phonetic contents and other information. This is good for speaker verification, but it will also limit the information that the attention component can attend to. On the other hand, the lower activations may be more noisy. It is a trade-off to choose a proper source to feed into the attention layer. In our experiments, we evaluate $\vec{k}_t$ with different $\vec{f}_t^l$. Multi-layer neural networks are also used to model the corresponding compatibility functions.

In addition, different features can be used as $\vec{a}_t$ in Eq (\ref{attention_3}).  In \cite{Zhang_attention_speaker}, bottleneck features extracted from a phonetically-aware neural network are served as the auxiliary features. Linguistic features extracted from text transcriptions can also provide useful information to the attention layer \cite{Wan_tts}. In this paper, we will only focus on the acoustic features and leave the auxiliary features to our future work.

The two attention variants in \cite{Chowdhury_attention_speaker}, cross-layer and divided-layer attention, are both special cases of the proposed framework. The cross-layer attention used the output of the 2nd-to-last layer as the key. The divided-layer attention used the same key as the cross-layer attention while used a 2-layer network as $\mathcal{G}(\cdot)$, whose first layer is a LSTM sharing the same structure with the last layer of the original network.

\subsection{Multi-head attention}

Multi-head attention is shown to be effective in machine translation \cite{attention_Vaswani}. Instead of performing a single attention function, multi-head attention uses a number of pooling layers to do attention in parallel. In the multi-head attention, the values, keys and query are first split to $h$ sub-vectors $\vec{v}^{(i)}$, $\vec{k}^{(i)}$ and $\vec{q}^{(i)}$ with $d_v'=d_v/h$, $d_k'=d_k/h$ and $d_q'=d_q/h$ dimensions, and $1\le i \le h$. These sub-vectors perform the attention pooling individually and then their results are concatenated. Unlike \cite{attention_Vaswani}, the concatenated outputs are not projected again, keeping the network architecture the same with the single head version. The multi-head attention can be denoted as
\begin{equation}\label{multihead_attention}
  \text{MultiHead}(\vec{v},\vec{k}, \vec{q}) = \text{Concat}(\text{Att}(\vec{v}^{(i)},\vec{k}^{(i)}, \vec{q}^{(i)}))
\end{equation}
In the multi-head attention, each head learns individual transformations and different heads can learn different sequence dependencies. From a view of model training, this allows the model to attend to information from different representation subspaces. Intuitively, it increases the modeling power of the single-head attention with no computational overhead.

\section{Experimental setup}

\subsection{Dataset}

We evaluate the performance of the attention mechanism on Fisher \cite{Cieri_fisher} and NIST SRE 2010 10s-10s datasets\cite{Martin_sre10}. The details of the two datasets are given as follows.
\begin{itemize}[leftmargin=*]
\item \emph{Fisher} is manually partitioned into training and evaluation subsets. We randomly select 878622 segments (sampled from 19386 utterances) as the training set. There are 9964 speakers in the 952h training set. The evaluation set contains 2000 speakers (1000 males and 1000 females) which do not overlap with the training set. The enrollment for each speaker consists of 2-8 segments, making the total duration ~10s for each speaker. The test data contains 6000 segments (3 segments per person and 1-2s per segment). The cross-gender trials are excluded, forming 6M target and non-target trials. Note that this dataset is larger and more difficult than the one used in our previous experiments. 
\item \emph{NIST SRE10} 10s-10s condition 5 are used in our experiments. NIST SRE 2004-2008 telephone excerpts, Switchboard Phase II Part 1/2/3 and Cellular Part 1/2 are used as the training set. This represents 5524 hours of data and comprises 6374 speakers from 64742 utterances.
\end{itemize}
All models in our experiments are gender-independent, and the results are reported on the male and female pooled trials. To fully illustrate the performance across different operation points, we report equal error rate (EER), minimum detection cost function from NIST SRE08 (minDCF08) \cite{Martin_sre08} and SRE10 (minDCF10) \cite{Martin_sre10} as the metrics.

\subsection{Baseline i-vector system}

An i-vector system is used as a complementary comparison to the standard and attention-based x-vector. 20-dimension static MFCCs with delta and delta-delta followed by cepstral mean normalization (CMN) are used as the acoustic features. Energy-based voice active detection (VAD) are applied. The features keep the same in our x-vector experiments.

The 2048-mixture universal background model (UBM) and the 600-dimension i-vector extractor are trained using the training data. LDA is applied to reduce the dimension of i-vector to 200 prior to PLDA scoring. On the Fisher dataset, LDA and PLDA is trained using a subset of the training set, which we found results in better performance.

\subsection{X-vector architecture}

Our x-vector architecture is kept the same with the one in Kaldi recipe SRE16 V2 \cite{Povey_kaldi}. The frame-level part of the x-vector network is a 5-layer time-delay neural network (TDNN) \cite{Peddinti_tdnn}. The input of each layer is the time-sliced output of the previous layer and the slicing parameter is: $\{t-2,t-1,t,t+1,t+2\}$, $\{t-2,t,t+2\}$, $\{t-3,t,t+3\}$, $\{t\}$, $\{t\}$.
Layer 1 to 4 have 512 nodes and the 5-th layer has 1500 nodes. The utterance-level part consists of a 2-layer fully-connected network with 512 nodes per layer. The final output is predicted by softmax and the size is the same as the number of speakers. 150-dimensionl LDA and PLDA scoring are trained and applied. Refer to \cite{Snyder_xvector} for more details.

\subsection{Attention-based pooling layer}

The attention-based pooling layer is applied to the x-vector architecture to replace the statistic average pooling layer. Activations from different layers are used as the keys and fully-connected networks are used as the compatibility function. All layers contain an affine transform followed by leaky ReLU and batch normalization. Different network settings are explored and will be described in the next section. For multi-head attention, the number of the heads is fixed at 50. The entire network is jointly optimized by cross entropy.

We implement all the neural networks using TensorFlow \cite{tensorflow}. Other pre- and post-processing are implemented with Kaldi toolkit. 

\section{Results}

\subsection{Fisher}
\begin{table}[thbp]
\caption{ Results on Fisher dataset. The attention layer using activations of the x-th layer as the key is denoted as att-x. The numbers in the brackets explicitly explain the network structure $\mathcal{G}(\cdot)$. For example, 100-500 is a 2-layer transform network with 100 and 500 nodes per layer. The att-5(500) represents the conventional attention layer used in many previous works. The number of the heads in the multi-head attention applied to att-4(500) is 50.
}
\label{table:result_fisher}
\centering
\begin{tabular}{|c|c|c|c|}
\hline
Systems & EER(\%) & minDCF08 & minDCF10 \\
\hline
\hline
i-vector & 10.35 & 0.0368 & 0.7863 \\
\hline
x-vector & 9.18 & 0.0325 & 0.8513 \\
\hline
\hline
att-5 (500) & 9.22 & 0.0325 & 0.8140 \\
\hline
att-4 (500) & 9.07 & 0.0320 & 0.7983 \\
\hline
att-3 (500) & 8.95 & 0.0325 & 0.8095 \\
\hline
\hline
att-4 (100-500) & 9.10 & 0.0325 & 0.8200 \\
\hline
att-3 (100-100-500) & 9.00 & 0.0324 & 0.8080 \\
\hline
\hline
att-4(500)+MultiHead & \textbf{8.91} & \textbf{0.0321} & \textbf{0.7835} \\
\hline
\end{tabular}
\end{table}

The baseline i-vector and x-vector are first compared on Fisher dataset. Table \ref{table:result_fisher} shows that x-vector with average pooling outperforms i-vector in EER and minDCF08, while i-vector performs better in minDCF10. We find that applying attention components generally improves the performance. As shown in this table, using activations from the 3rd and 4th layer as the keys achieves better results than using output from the last layer. However, it seems that using a multi-layer network to model the key transformation does not bring any benefits. We guess the reason is that the number of nodes in these transform networks is too small and cannot fully capture the characteristics of the keys. Finally, the multi-head attention achieves the best performance, which illustrates the effectiveness of this method.

\begin{figure}[t]
    \centering
    \includegraphics[width=1.0\linewidth]{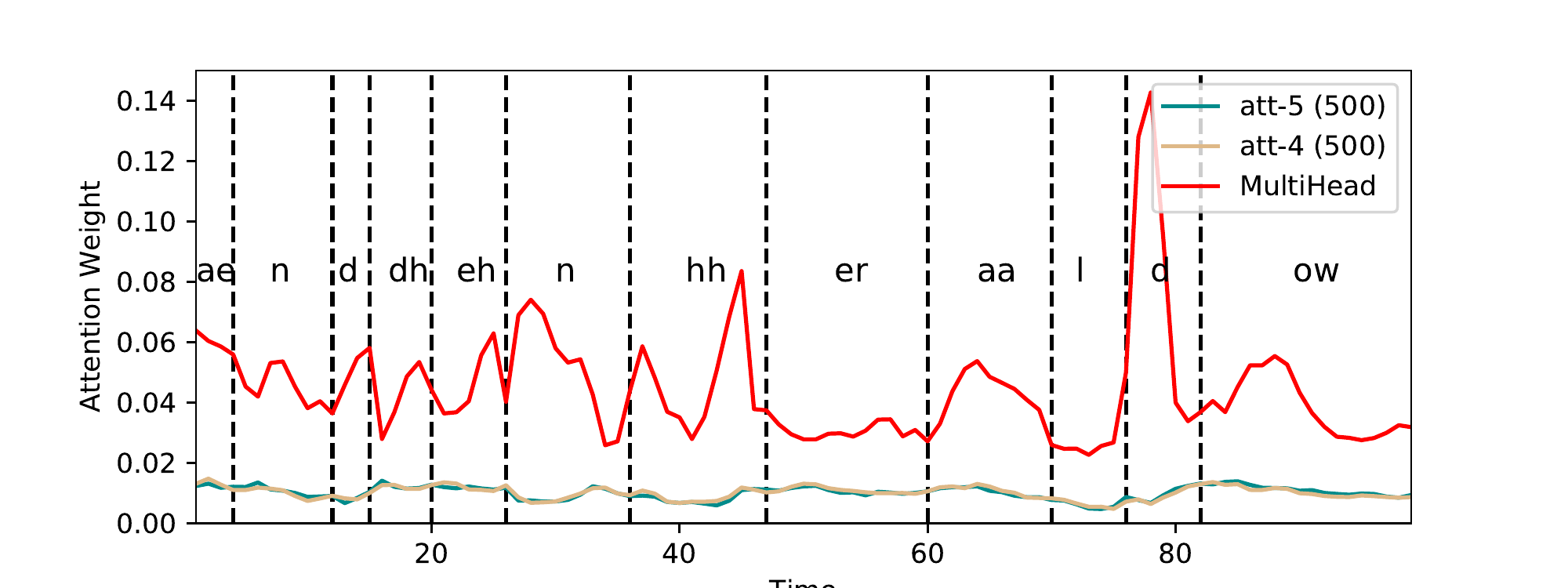}
    \caption{The trajectory of the attention weights extracted from different attention layers. The weights from the multi-head attention are always the largest. The utterance is randomly sampled from the test set.}
    \label{fig:attention}
\end{figure}

Then, we explore the attention weights extracted from different attention layers. For multi-head attention, we plot the maximum weights at each time stamp. The results are presented in Fig. \ref{fig:attention}. This figure shows that the attention weights calculated from the outputs of the 4-th and 5-th layers are quite similar.  The weight distribution across the entire utterance is more flat than we expected. The voiced regions do not get much larger weights than the unvoiced regions. This is interesting and worth further investigation. 

Although the weights calculated from the 4-th and 5-th layers are similar, the former one still results in better performance in Table \ref{table:result_fisher}. As stated before, the final output of the frame-level network may suppress some information which is not directly related to speakers.  However, this auxiliary information, as we discussed in the previous section, can be useful to guide the attention layer from different feature spaces. In this experiment, using the output of a lower layer as the key is a better choice.

For the multi-head attention, we find that the maximum weights are always much larger than the single-head versions. Since these weights come from different heads, the phenomenon indicates that the weights in each head are spiky at different positions. By this way, the multi-head attention can attend to different information of the utterance.

\subsection{NIST SRE10}

\begin{table}[thbp]
    \caption{ Results on NIST SRE10 10s-10s condition 5. The abbreviations in this table follow the same manner as Table \ref{table:result_fisher}. }
    \label{table:result_sre}
    \centering
    \begin{tabular}{|c|c|c|c|}
        \hline
        Systems & EER(\%) & minDCF08 & minDCF10 \\
        \hline
        \hline
        i-vector & 10.46 & 0.0533 & 0.9817 \\
        \hline
        x-vector & 10.81 & 0.0530 & 0.9304 \\
        \hline
        \hline
        att-5 (500) & 10.44 & 0.0497 & 0.9469 \\
        \hline
        att-4 (500) & 10.25 & 0.0496 & 0.9707 \\
        \hline
        att-3 (500) & 10.85 & 0.0510 & 0.9652 \\
        \hline
        \hline
        att-4 (100-500) & 10.99 & 0.0498 & 0.9047 \\
        \hline
        att-3 (100-100-500) & 10.25 & 0.0493 & \textbf{0.8663} \\
        \hline
        \hline
        att-4(500)+MultiHead & \textbf{9.67} & \textbf{0.0486} & 0.9111 \\
        \hline
    \end{tabular}
\end{table}

The performance of our attention framework is also evaluated on the NIST SRE10 10s-10s condition 5. The x-vector is slightly worse than the conventional i-vector in EER and performs better in minDCF08 and minDCF10. The conventional attention layer, \emph{att-5(500)}, improve the performance in EER and minDCF08, while the performance in minDCF10 is slightly worse. The performance is similar when using different sources as the attention key. 
Using the multi-head attention improves the results across all these operation points and outperforms the conventional x-vector by 10\%, 8\% and 2\% in EER, minDCF08 and minDCF10, respectively.

\section{Conclusions}

In this paper, we first propose a unified attention framework for speaker verification. Unlike previous works, outputs from different hidden layers are used as the key of the attention-based pooling component. We further introduce the multi-head attention into this framework. The multi-head attention is able to explore information from different feature subspaces and provides better performance.
The speaker embeddings extracted using single and multi-head attention pooling both outperform the average pooling based x-vector on the Fisher dataset. 
The experiment also shows that the attention weights computed from the activations of lower layers can result in better performance. On NIST SRE10 10s-10s, most of the single-head attention outperforms the baseline in EER and minDCF08, while the improvement in minDCF10 is inconsistent. With multi-head attention, the proposed method achieves a better result than the conventional x-vector across all these operation points. In our experiments, using multi-layer fully-connected networks to model the key transformation does not necessarily bring much benefits.

In the future, we will explore the use of the auxiliary features. Bottleneck features extracted from an ASR network and linguistic features can be the candidates to improve the performance of the attention mechanism.

\section{Acknowledgements}
The work is supported by National Natural Science Foundation of China under Grant No. 61370034, No. 61403224 and No. 61273268.

\bibliographystyle{IEEEtran}
\bibliography{mybib}

\end{document}